\documentclass[aps,floatfix,prd,nofootinbib,superscriptaddress,reprint,showpacs,10pt,preprintnumbers,longbibliography,showdoi]{revtex4-1}
\usepackage[utf8]{inputenc}
\usepackage[pdftex]{graphicx}
\usepackage{float}
\usepackage{amsmath}
\usepackage{braket}
\usepackage{amssymb}
\usepackage{mathtools}
\usepackage{siunitx}
\usepackage{amsfonts}
\usepackage{dsfont}
\usepackage{array}
\usepackage{bm}
\usepackage{mathrsfs}
\usepackage{pifont}
\usepackage{multirow}
\usepackage{upgreek}
\usepackage[dvipsnames]{xcolor}
\usepackage{changes}
\usepackage[pdftex,
  pdftitle={Canonical Momenta in digitized SU(2) Lattice Gauge Theory:
  Definition and Free Theory},
  pdfauthor={M. Garofalo, T. Hartung, T. Jakobs, K. Jansen,
    J. Ostmeyer, S. Romiti and C. Urbach},
  bookmarks,
  colorlinks,
  linkcolor=myblue,
  citecolor=mymagenta,
  menucolor=black,
  urlcolor=myblue,
  plainpages=false,
  pdfpagelabels,
  hypertexnames=false]{hyperref}
\usepackage{verbatim}
\usepackage{slashed}
\usepackage{cleveref}
\usepackage{ucs}
\usepackage{subfigure}

\usepackage{tikz}

\crefformat{figure}{FIG.~#2#1#3}
\crefformat{equation}{Eq.~(#2#1#3)}

\newcommand{\ri}{\mathrm{i}}
\newcommand{\SUTwo}{SU$(2)$}
\newcommand{\abs}[1]{|#1|}
\newcommand{\matr}[2]{\left(\begin{array}{#1}#2\end{array}\right)}

\definecolor{mymagenta}{RGB}{200, 0, 100}
\definecolor{myblue}{RGB}{45, 48, 146}

\definecolor{THblue}{rgb}{0.47, 0.62, 0.8}

\usepackage[normalem]{ulem}

\graphicspath{{plots/}}

\begin{document}
\title{Canonical Momenta in Digitized SU$(2)$ Lattice Gauge Theory:\\ Definition
  and Free Theory}

\author{Timo Jakobs}
\affiliation{Helmholtz-Institut für Strahlen- und Kernphysik, University of Bonn, Nussallee 14-16, 53115 Bonn, Germany}
\affiliation{Bethe Center for Theoretical Physics, University of Bonn, Nussallee 12, 53115 Bonn, Germany}

\author{Marco Garofalo}
\affiliation{Helmholtz-Institut für Strahlen- und Kernphysik, University of Bonn, Nussallee 14-16, 53115 Bonn, Germany}
\affiliation{Bethe Center for Theoretical Physics, University of Bonn, Nussallee 12, 53115 Bonn, Germany}

\author{Tobias Hartung}
\affiliation{Northeastern University - London, Devon House, St Katharine Docks, London, E1W 1LP, UK}

\author{Karl Jansen}
\affiliation{CQTA, DESY Zeuthen, Platanenallee 6, 15738 Zeuthen, Germany}
\affiliation{Computation-Based Science and Technology Research Center,
  The Cyprus Institute, 20 Kavafi Street, 2121 Nicosia, Cyprus}

\author{Johann Ostmeyer}
\affiliation{Department of Mathematical Sciences, University of Liverpool, Liverpool, L69 7ZL, United Kingdom}

\author{Dominik Rolfes}
\affiliation{Helmholtz-Institut für Strahlen- und Kernphysik, University of Bonn, Nussallee 14-16, 53115 Bonn, Germany}
\affiliation{Bethe Center for Theoretical Physics, University of Bonn, Nussallee 12, 53115 Bonn, Germany}

\author{Simone Romiti}
\affiliation{Helmholtz-Institut für Strahlen- und Kernphysik, University of Bonn, Nussallee 14-16, 53115 Bonn, Germany}
\affiliation{Bethe Center for Theoretical Physics, University of Bonn, Nussallee 12, 53115 Bonn, Germany}

\author{Carsten Urbach}
\affiliation{Helmholtz-Institut für Strahlen- und Kernphysik, University of Bonn, Nussallee 14-16, 53115 Bonn, Germany}
\affiliation{Bethe Center for Theoretical Physics, University of Bonn, Nussallee 12, 53115 Bonn, Germany}

\date{\today}
\begin{abstract}
  
Hamiltonian simulations of quantum systems require a
finite-dimensional representation of the operators acting on the
Hilbert space $\mathcal{H}$. 
Here we give a prescription for gauge links and canonical momenta of
an SU$(2)$ gauge theory, such that the matrix representation of the
former is diagonal in $\mathcal{H}$. 
This is achieved by discretising the sphere $S_3$ isomorphic to SU$(2)$
and the corresponding directional derivatives.
We show that the fundamental commutation relations are fulfilled up to
discretisation artefacts.
Moreover, we directly construct the Casimir operator corresponding to
the Laplace-Beltrami operator on $S_3$ and show that the spectrum of
the free theory is reproduced again up to discretisation effects.
Qualitatively, these results do not depend on the specific
discretisation of SU$(2)$, but the actual convergence rates do.

 \end{abstract}

\maketitle

\section{Introduction}

While the Hamiltonian of lattice gauge theories is
known since 1975~\cite{Kogut:1974ag}, it only recently received a fresh
interest. This is due to the development of tensor network state (TNS)
and quantum computing (QC) methods over the last few years. These
methods promise to provide the possibility to investigate lattice
gauge theories (and other quantum systems) in regions of parameter
space inaccessible to Monte Carlo methods, most prominently in
situations when a sign problem is hindering the application of
stochastic methods. In addition, with the Hamiltonian formulation of 
lattice field theories also real time simulations become possible, 
opening up new insights in the dynamical properties of physical systems. 

Tensor network based methods have been introduced for lattice field
theory methods in many works, see,
e.g.\ Refs.~\cite{Silvi2014,Dalmonte2016,Banuls2019SimulatingLG,Banuls2019,Banuls2020TNreview}. The
success of TNS lies on the fact  
that only a small subspace of the complete Hilbert space describes the
physically often only relevant low energy physics. Therefore, 
various phenomena such as string breaking and real-time 
dynamics~\cite{Buyens2013,Kuehn2015,Pichler2016,Buyens2016b,Banuls2019b,Rigobello2021} 
or the phase structure of gauge theories at finite fermionic densities~\cite{Banuls2016a,Silvi2017,Felser2019,Silvi2019} 
have been investigated using TNS on moderately large lattice volumes.

More recently, quantum computer simulations have been performed for lattice gauge theories. In this approach, 
the number of required qubits grows only linearly with the number of lattice sites. 
There are also proposals to implement real-time dynamics for scalar quantum field theories and 
quantum electrodynamics~\cite{Byrnes2006, Jordan2012, Mathis2020}. 
Since quantum computations use the Hamiltonian formulation, they can completely avoid the sign problem. 
Quantum computers therefore allow to realise Feynman's vision to simulate nature on 
a quantum mechanical, physical system~\cite{Feynman1982}. 

The literature of quantum computations for lattice gauge theories has 
increased tremendously in the last years, see Refs.~\cite{Banuls2019SimulatingLG,Banuls2019,Klco2019,Atas2021,Ciavarella2021,Clemente:2022cka}. 
There are various 
approaches for implementing lattice gauge theories using optical
lattices~\cite{Banerjee2012,Tagliacozzo2013,Tagliacozzo2013a}, atomic
and ultra-cold quantum
matter~\cite{Buchler2005,Zohar2011,Zohar2012,Hauke2013,Zohar2013a,Zohar2013c,Banerjee2013,Zohar2015a,Laflamme2015,Gonzalez-Cuadra2017,Rico2018,Zache2018},
and further proof-of-principle implementations on a real
superconducting
architecture~\cite{Klco2018,Klco2019,Atas2021,Ciavarella2021,Mazzola2021}
and real-time and variational simulations on a trapped ion
system~\cite{Martinez2016, Kokail2019}. For recent overviews
see Refs.~\cite{Dalmonte2016,Banuls2019SimulatingLG,Banuls2019,Funcke:2023jbq}. 

A very important aspect of quantum computations is the quest for the most efficient
discretisation scheme of the corresponding gauge group needed to apply
both, TNS and QC methods, see
e.g.\ Ref.~\cite{Davoudi:2020yln}. Work in this direction, both for Abelian 
gauge theories 
with and without fermions has been performed by a number of groups already, see
for
instance Refs.~\cite{Klco:2018kyo,Lewis:2019wfx,Paulson:2020zjd,Haase:2020kaj,Nguyen:2021hyk,Clemente:2022cka}.
Also for non-Abelian SU$(2)$ and SU$(3)$ lattice gauge theories there
are a number of works available~\cite{Klco:2019evd,Atas:2021ext,ARahman:2021ktn,Ciavarella:2021nmj,ARahman:2022tkr,Atas:2022dqm,Catumba:2022ced,ARahman:2022jks,Gustafson:2022xdt,Farrell:2022wyt,Farrell:2022vyh}.
For more algorithmic developments we refer to
Refs.~\cite{Alam:2021uuq,Carena:2022kpg,Gustafson:2023swx}.
Another possible formulation is provided
by the so-called quantum link
model~\cite{Chandrasekharan:1996ih,Brower:1997ha,Wiese:2021djl}.

In particular for non-Abelian SU$(N_c)$ lattice gauge theories it is important
to understand how to most efficiently digitise the gauge field operators
$\hat U$ and the corresponding canonical momentum operators $\hat L$ and $\hat R$
constituting the Hamiltonian
\begin{equation}
  \label{eq:hamiltonian}
  \hat H =\ \frac{g_0^2}{4}\sum_{\mathbf{x},c,k} \left(\hat{L}_{c,k}^2(\mathbf{x}) +
  \hat{R}_{c,k}^2(\mathbf{x})\right) - \frac{1}{2g_0^2}\sum_{\mathbf{x},k<l} \mathrm{Tr}\,
  \mathrm{Re}\, \hat{P}_{kl}(\mathbf{x})\,.
\end{equation}
In the above sums, $\mathbf{x}$ represent the coordinates of a $d$-dimensional lattice
and $k,l$ label the corresponding directions with $\hat l, \hat k$ unit
vectors in these directions.
$c = 1, \ldots, N_c^2-1$ indexes the generators of the algebra,
$g_0$ is the (bare) gauge coupling constant and the plaquette
operator is defined as
\begin{equation}
  \label{eq:plaquette}
  \hat{P}_{kl}(\mathbf{x})\ =\ \hat{U}_k(\mathbf{x})\, \hat{U}_l(\mathbf{x}+\hat k)\,
  \hat{U}^\dagger_k(\mathbf{x}+\hat l)\, \hat{U}^\dagger_l(\mathbf{x})\,.
\end{equation}
The $\hat{U}$ are the gauge field operators as explained further below
and the trace is taken in colour space.

In the literature the first and second sum in \cref{eq:hamiltonian} 
are called respectively \mbox{(chromo-)}electric and \mbox{(chromo-)}magnetic part.
Most of the investigations of non-Abelian lattice gauge theories in
the Hamiltonian formalism chose a basis of the Hilbert space
$\mathcal{H}$ such that the electric part is diagonal. This leads for
instance to the so-called character expansion or loop-string
formulations, the status of which is elaborately discussed in
Ref.~\cite{Davoudi:2022xmb}. 

In this paper we are going to explore a different pathway: following the
ideas discussed for U$(1)$ in Refs.~\cite{Haase:2020kaj,Clemente:2022cka} we will
develop a formulation with a basis of $\mathcal{H}$ in which the
non-Abelian gauge field operators are diagonal. For this purpose we
use the digitisations we recently proposed in
Ref.~\cite{Hartung:2022hoz,Jakobs:2022ugr}, which provide a natural discretised
parametrization of SU$(2)$ and which can be extended to larger
$N_c$-values. For more works related to digitised SU$(N_c)$ gauge fields see
Refs.~\cite{Ji:2020kjk,Alexandru:2019nsa,Alexandru:2021jpm,Ji:2022qvr}. 

The remaining task is to find the corresponding digitised versions of
the operators $\hat L, \hat R$ or directly the Casimir operator $\hat
L^2 + \hat R^2$, 
which we are going to discuss in the following. We will show that it
is possible to find discrete versions of $\hat L, \hat R$ fulfilling
the fundamental commutation relations up to discretisation
effects. Moreover, we show that in order to reproduce the spectrum of
the free Hamiltonian, the aforementioned Casimir operator needs to be
discretised directly. If this is done, spectrum and eigenstates of the
free Hamiltonian are reproduced up to discretisation effects, the size
of which depends on the specific choice of the partitioning of SU$(2)$.

\section{Commutators and State Space}

To be concrete, we define the set of coordinates of the
$d$-dimensional spatial lattice as
\[
\Lambda\ =\ \{\mathbf{x}\in\mathbb{R}^d: x_k = 0, a, 2a, \ldots, (L_k-1)a\}\,,
\]
with $k$ labelling the directions as above.
$a$ is the lattice spacing which we set to $a=1$ in the following and
$L_k\in\mathbb{N}$ the lattice extent in direction $k$.
The quantization conditions are imposed at each point of the space time lattice.
This gives freedom for any choice of the boundary conditions, 
which are inessential for the rest of our discussion.

States on the full lattice are constructed by tensor products
of basis states for each lattice site and direction. This is why it is
sufficient to discuss the discretisation for one lattice site and
direction and, thus, we will drop the spatial coordinates $\mathbf{x}$
and the directions $k$. 

Classically, for each lattice site and direction the gauge link $U$ is
an SU$(N_c)$ matrix in the fundamental representation,  with $N_c^2$
elements $u_{ij}$. 
On the quantum level, the elements of the gauge fields
$\hat u_{ij}\in\mathbb{L}$ become operators, with $\mathbb{L}$ the
linear operators $\mathcal{H}\to\mathcal{H}$. Now, $\hat{U}$ is a
$N_c\times N_c$ operator valued matrix, which is constrained as follows:
given a gauge field state $|U\rangle\in\mathcal{H}$ of the system with
$\hat{U}|U\rangle = U|U\rangle$ we require
that $U\in\mathrm{SU}(N_c)$ in the fundamental representation.
$\hat{L}_{c},\hat{R}_c\in\mathbb{L}$ on the other hand represent 
the corresponding canonical momenta (in the adjoint representation),
which are generating the left and right gauge transformations.

Given the $\hat{U}$, the momenta are defined via the fundamental
commutation relations 
\begin{equation}
  \label{eq:commUL}
        [\hat{L}_c, \hat{U}_{mn}]\ =\ (t_c)_{mj}\, \hat{U}_{jn}\,,
        \qquad [\hat{R}_c, \hat{U}_{mn}]\ =\ \hat{U}_{mj}(t_c)_{jn}\,.
\end{equation}
Here, $t_c$ are the generators of the corresponding Lie algebra.
Moreover, the $\hat{L}_c$ resemble the group structure
\begin{equation}
  \label{eq:commLL}
        [\hat{L}_a, \hat{L}_b] = f_{abc} \hat{L}_c\,,
\end{equation}
with the the structure constants $f_{abc}$ of the 
algebra, and likewise the $\hat{R}_c$.

Specifically, for SU$(2)$, the generators are given by the Pauli
matrices $t_c$ with indices $c=1,2,3$. 
We parametrise the basis for the gauge field states as follows:
$U\in$SU$(2)$ can be parametrised with three real valued parameters
$y_0, y_1, y_2$ as follows
\begin{equation}
  \label{eq:isomorphy}
  U = \begin{pmatrix}
    y_0 + \ri y_1 & y_2 + \ri y_3 \\
    -y_2 + \ri y_3 & y_0 - \ri y_1\\
  \end{pmatrix}\,,\quad y_3^2 = 1- \sum_{i=0}^2 y_i^2\,.
\end{equation}
Since SU$(2)$ is isomorphic to the sphere $S_3$, we can also write
$y=(y_0, y_1, y_2, y_3)^t \in S_3$. 
Accordingly, we define operators $\hat y_j:\mathcal{H}\to\mathcal{H}$
by the following action 
\[
\hat y_j |U(y)\rangle\ =\ y_j |U(y)\rangle\,,\quad j=0,1,2,3\,,
\]
and
\[
\begin{split}
  \hat u_{00} = \hat y_0 + \ri \hat y_1\,,\quad \hat u_{01} = \hat y_2
  + \ri \hat y_3\,,\\
  \hat u_{10} = -\hat y_2 + \ri \hat y_3\,,\quad \hat u_{11} = \hat y_0
  - \ri \hat y_1\,.\\
\end{split}
\]
This defines the action of $\hat U:\mathcal{H}\to\mathcal{H}$ on a
given state via
\[
\hat U =
\begin{pmatrix}
  \hat u_{00} & \hat u_{01}\\
  \hat u_{10} & \hat u_{11}\\
\end{pmatrix}\,.
\]
Therefore, the $y_{0,1,2}$ can be regarded as quantum numbers
labelling the states $|y_0, y_1, y_2\rangle \equiv |U(y)\rangle$ which are simultaneous eigenstates of
operators $\hat y_{0,1,2}$.

Alternatively, one could also work with three angles $\vec\alpha$
which can be used to parametrise $U = \exp(\ri\, \vec\alpha\cdot\vec
t)$. Given $\vec\alpha$, the $y_i$ can be readily computed. For
quantisation, each angle $\alpha_i$ is then promoted to a linear operator
$\mathcal{H}\to\mathcal{H}$.

Formally, the canonical momenta are defined as Lie derivatives:
\begin{equation}
  \label{eq:momenta}
  \begin{split}
    \hat{L}_c\, f(U)\ &= -\ri \frac{\mathrm{d}}{\mathrm{d}\beta}\,
    f\left(e^{\ri\,\beta t_c}\, U\right)|_{\beta = 0}\,,\\\qquad
    \hat{R}_c\, f(U)\ &= -\ri \frac{\mathrm{d}}{\mathrm{d}\beta}\,
    f\left(U\, e^{\ri\,\beta t_c}\right)|_{\beta = 0}
  \end{split}
\end{equation}
for a function $f(U)$, $f:\mathrm{SU}(2) \rightarrow \mathbb{R}$. 

The states parametrised by $y_0, y_1$ and $y_2$ can be discretised
using one of the partitionings we proposed in
Ref.~\cite{Hartung:2022hoz}, or any other partitioning of SU$(2)$.
In fact, the precise form of the partitioning becomes only relevant
for our numerical experiments presented in later sections.

Next we will discuss how to discretise the momenta \cref{eq:momenta}
for such a partitioning. A first approach to this problem can be found
in Ref.~\cite{Garofalo:2022swa}. However, while there we could define
the momenta such that the fundamental commutation relations are
fulfilled up to discretisation effects, the na\"ive approach of
squaring the so constructed operators does not lead to a free
Hamiltonian for which the spectrum converges.

\section{Construction of Triangulated Derivatives}
\label{sec:triangulation}

We consider an arbitrary finite subset $\mathcal{D} = \{ D_i \}
\subset \mathrm{SU}(2)$. We will now discuss how to construct $\hat L,
\hat R$ based on finite
element methods, developed for triangulated manifolds.
For a more in depth introduction to this type of methods we recommend 
Refs.~\cite{Crane:2013:DGP,correa5262940,Crane2019TheNC}.
To make use of these methods, we again employ the isomorphism between
\SUTwo\ and $S_3$: 
$L_c$ and $R_c$ can be also understood as covariant derivatives on
$S_3$ in directions 
\begin{equation}
  v_{L_c} = t_c \, U \qquad \text{and} \qquad  v_{R_c} =  U \, t_c
\end{equation}
at point $U$.
Furthermore $\{ t_c \, U \, | \, c \in \{1,2,3\} \}$ forms an orthonormal basis
of the tangent space at point $U$. The same holds for $\{ U  \, t_c  \, | \, c \in \{1,2,3\} \}$.
This means that for the continuous version of the operators
\begin{equation}
	\sum_c \hat{L}^2_c = \sum_c \hat{R}^2_c = -\Delta\,,
\end{equation}
where $\Delta$ denotes the Laplace-Beltrami operator on $S_3$.

In the following we will construct the covariant derivative as well
as the Laplace-Beltrami operator following the methods presented in
Ref.~\cite{correa5262940} and Ref.~\cite{Crane2019TheNC}, respectively.

\subsection{Construction of the Gradient}\label{subsec:construction-gradient}

To construct a discrete version of the covariant derivative, we first
need to perform a Delaunay triangulation~\cite{Delaunay:1934} of the
points on $S_3$ embedded in $\mathbb{R}^4$. This triangulation connects
the points of our partitioning to 3-simplices (tetrahedons), such that
the ball spanned by the vertices of each simplex does not contain any
of the other vertices. The result is a set of simplices
\begin{equation}
  \mathcal{C} = \left\{ (i_0, i_1, i_2, i_3) \right\}\,,
\end{equation}
where $i_0, \dots, i_3$ label the vertices $D_{i_k}$ of each simplex.
In our case in four dimensions the simplices are tetrahedrons, i.e. they
are built from four vertices. 

Next we discretise the functions by introducing the basis functions
on $S_3$ which have to property
\begin{equation}\label{eq:basis-hat}
 \phi_j(D_i) =\phi_j(i) = \delta_{ij}
\end{equation}
on the vertices of our partitioning, while within a simplex connected
to vertex $j$ they are
linear piece-wise functions that interpolate the values between the
corresponding vertices. Within all other, not connected simplices
$\phi_j$ is identically zero.
With these definitions we can construct
a discrete version $\tilde{f}$ of an arbitrary function $f: \mathrm{SU}(2) \rightarrow \mathbb{R}$
\begin{align}
  \tilde{f}(i)  = & \sum_j f(D_j) \,\phi_j(i)\,.
\end{align}
For the linear interpolation, we introduce local coordinates
$\vec\alpha$ relative to vertex $i_0$ by noting that every $U$ in a
simplex $C\in\mathcal{C}$ can be written as
\begin{equation}
  U=\exp(\ri \vec{\alpha} \cdot \vec{t})D_{i_0}\,.
  \label{eq:lcLocalCoords}
\end{equation}

\begin{figure}
  \centering
  \includegraphics{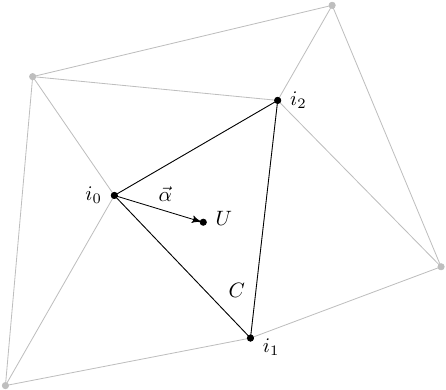}
  \caption{We visualise the construction of the local coordinates
    $\vec{\alpha}$ of a gauge group element $U \notin \mathcal{D}$
    found in 2-simplex (a triangle) $C$ of the triangulation for the simpler two
    dimensional case. $i_0$ is arbitrarily chosen as the origin of 
    the local coordinate system. From there we can then calculate the
    rotation angles $\vec{\alpha}$ by using
    \cref{eq:lcLocalCoords}.}
  \label{fig:simplexCoords}
\end{figure}

\begin{figure}
  \centering
  \includegraphics{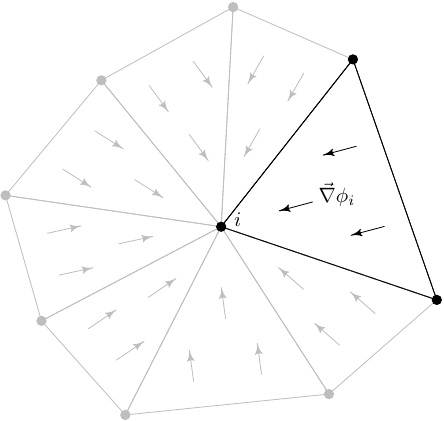}
  \caption{The direction $\vec{\nabla} \phi_i$ in the simplices connected to vertex $i$ for a two dimensional example. As can be seen the  covariant derivative is orthogonal to the side of the triangle opposing $i$. Its magnitude will be the inverse of the distance to said side.}
  \label{fig:simplexGrad}
\end{figure}

A sketch of this construction for the two dimensional case can be
found in \cref{fig:simplexCoords}.
We can then use $\vec\alpha$ to approximate $f(U)$ within $C$ as
\begin{equation}
  f(U) = \tilde{f}(i_0) + \vec{\nabla} f_C \cdot \vec{\alpha} + \mathcal{O}(\alpha^2)\,,
  \label{eq:cell-gradient-red}
\end{equation}
with $\vec \nabla f_C$ corresponding to the covariant derivative in
$C$ we want to compute.
In order to reproduce the function values at the vertices $i_{1,2,3}$
with local coordinates $\vec\alpha_{1,2,3}$, $\vec{\nabla} f_C $ needs to
fulfil the linear equation
\begin{equation}
  \begin{pmatrix}
    \vec{\alpha}^T_1 \\
    \vec{\alpha}^T_2 \\
    \vec{\alpha}^T_3
  \end{pmatrix} \vec\nabla f_C = \begin{pmatrix}
    \tilde{f}(i_1) - \tilde{f}(i_0) \\
    \tilde{f}(i_2) - \tilde{f}(i_0) \\
    \tilde{f}(i_3) - \tilde{f}(i_0)
  \end{pmatrix}.
\end{equation}
For the special case of the basis function $\tilde f = \phi_i$ 
only simplices containing $i$ will have a non trivial covariant
derivative. As seen in \cref{fig:simplexGrad} (again for the two
dimensional case) it will be pointing in
the direction of the normal vector of the face opposite to the vertex
$i$. 

Similarly, for $R_c$ a right $\vec \nabla$ can be constructed by
introducing coordinates from
\begin{equation}
  U = D_{i_0} \exp\left( \ri \vec{\alpha} \cdot \vec{t} \, \right)
  \label{eq:rcLocalCoords}
\end{equation}
instead of \cref{eq:lcLocalCoords} followed by the same steps as above.

A good approximation of the covariant derivative at a vertex $i$ can be
obtained by taking a weighted average over all simplices $C$ which
have $i$ as one of their vertices:
\begin{equation}
  \tilde{ \nabla} {f} (i) = \frac{1}{W_i} \sum_{\{C \in \, \mathcal{C}
    | i \in C \}} w_C \vec \nabla f_C 
\end{equation}
with weights $w_C$ and
\begin{equation}
  W_i = \sum_{\{C \in \, \mathcal{C} | i \in C \}} w_C\,.
\end{equation}
The weight $w_C$ can be calculated in different ways as e.g. presented in Ref.~\cite{correa5262940}. We consider here
weighting by the simplex volume
\begin{equation}
  w^{\text{vol}}_{C} = \mathrm{Vol}(C) = \frac{1}{6} \left| \det\left(\vec{\alpha}_1 \, \vec{\alpha}_2 \,\vec{\alpha}_3  \right) \right| \,.
\end{equation}
Numerical experiments show that this choice of weights works well. Therefore, we leave investigations of alternative choices for future work. It might regain importance in conjunction with ensuring Gauss' law.

The matrix elements of the operator $\hat L_c$ are then calculated as
\begin{equation}
  \hat{\vec{L}}_{ij} = 
  \begin{pmatrix}
    \hat{L}_{1\, ij} \\
    \hat{L}_{2\, ij} \\
    \hat{L}_{3\, ij}
  \end{pmatrix} =- \ri \tilde\nabla\phi_j (i)\,.
\end{equation}
The same holds for $R_c$, the only difference being the different calculation
of the local coordinates in \cref{eq:rcLocalCoords}.

\subsection{Construction of the Laplace-Beltrami Operator}

As our approximation of the covariant derivative is based on linear interpolation,
simply calculating $\sum_c L_c^2$ is unlikely to give good results for the Laplace-Beltrami operator.
It requires second order derivatives which are intrinsically ignored in linear approximations.
Instead, a common approach is to approximate the operator by making use of Greens identity.

We start from the Laplace equation
\begin{equation}
  \Delta u =f\label{eq:laplace-eq}
\end{equation}
and we first introduce the inner product
\begin{equation}
  \begin{split}
    \langle f, g \rangle \coloneqq  \sum_{\{C \in \mathcal C\}} \int_{C} \mathrm{d}V \, & f(\exp\left( \ri\vec{\alpha} \cdot \vec{t} \, \right) U_{i_0}) \\
    & \qquad g(\exp\left( \ri\vec{\alpha} \cdot \vec{t} \, \right) U_{i_0})
  \end{split}\label{eq:scalar-product}
\end{equation}
where the integral is carried out over the local coordinates defined
in \cref{eq:lcLocalCoords} and the integral over the volume is split
into the sum of the integrals over simplices. We can project \cref{eq:laplace-eq} to a basis function 
$\phi$ defined in \cref{eq:basis-hat}
\begin{equation}
  \label{eq:laplace-eq-proj}
  \langle \Delta u, \phi_i\rangle =\langle f,\phi_i\rangle\,.
\end{equation}
The left hand side can be computed using Green's theorem
\begin{equation}
  \begin{split}
    \langle \Delta u, \phi_i \rangle &= \sum_{\{C \in \mathcal C\}}\int_{C} \mathrm{d}V (\Delta u) \phi_i \\
    &=- \sum_{\{C \in \mathcal C\}}\int_{C} \mathrm{d}V (\vec\nabla u) \cdot 
    (\vec\nabla \phi_i )\\
    &\quad+\sum_{\{C \in \mathcal C\}}\int_{\partial C} \mathrm{d}S\, \vec{n}
    \cdot (\vec\nabla u)   \phi_i \,.
  \end{split}
\end{equation}
Here $\vec{n}$ denotes the normal vector of the simplex $C$. As the normal vectors of the two simplices connected at a given face will oppose each other, the boundary term vanishes, when summing over all simplices. In the next step we are approximating the function $u$ with its projection on the space spanned by the basis functions, i.e. $\tilde u=\sum_j u_j \phi_j$ obtaining the matrix equation
\begin{equation}\label{eq:laplacian-lhs-M}
  \begin{split}
    &\langle \Delta u, \phi_i \rangle \approx-
    \sum_j u_j \sum_{\{C \in \mathcal C\}}\int_{C} \mathrm{d}V (\vec\nabla \phi_j) \cdot 
    (\vec\nabla \phi_i )= S_{ij} u_j	\,.
  \end{split}
\end{equation}
The gradient of the basis function inside a simplex is a constant and the integral over the volume can be computed explicitly
\begin{equation}
  S_{ij}=-\sum_{\{C \in \mathcal C| i,j \in C\}}(\vec\nabla \phi_j) \cdot 
  (\vec\nabla \phi_i )\mathrm{Vol} (C)\,.
\end{equation}
We approximate the integral on the right-hand side of \cref{eq:scalar-product}
as a Riemann sum
\begin{equation}
  \label{eq:laplacian-rhs-M}
  \langle f,\phi_i\rangle \approx \sum_j f(j) \, v(j) \, \phi_i (j)  = f(i) \, v(i)\,.
\end{equation}
As weights we will use the volumes of the cells of the barycentric dual of the
triangulation. These distribute the volume of each simplex equally onto each of
its vertices and are thus given by
\begin{equation}
  \label{eq:volbary}
  v(i)=\sum_{\{C \in \mathcal C | i \in C\}} \frac{\mathrm{Vol}(C)}{4}\,.
\end{equation}
Equating \cref{eq:laplacian-lhs-M,eq:laplacian-rhs-M} one obtains
\begin{equation}
  S_{ij} u_j = v_i f_i\,.
\end{equation}
Thus, the matrix form of $\hat{\vec{L}}^2$ can then be calculated as
\begin{equation}
  \label{eq:Lsq}
  \hat{\vec{L}}^2_{ij}=-  \frac{1}{v_i}S_{ij}.
\end{equation}
As the Laplacian is independent of local coordinates,
$\hat{\vec{R}}^2=\hat{\vec{L}}^2$ holds like
in the continuum also for the discretised operators.
Note that like for the linear operator $\hat{\vec L}_{ij}$, the matrix
form of the discretised operator $\hat{\vec{L}}^2_{ij}$ is still local
and sparse, even though we have used a global integral definition for
its construction. This will no longer be the case if a
higher order integration scheme is adopted in
\cref{eq:laplacian-rhs-M}, in which case one has to take a matrix $M$
on the right had side of the discretised \cref{eq:laplace-eq} into
account, i.e. $S_{ij}u_j=M_{ij}f_j$. As a consequence one obtains
$\hat{\vec{L}}_{ij}^2=-M^{-1}_{ik}S_{kj}$ with  $M^{-1}$ a dense
matrix.

\subsection{Partitionings of SU$(2)$}

For convenience we compile here the definitions for the partitionings
of SU$(2)$ we are going to use in the following. More details on the
first four partitionings can be found in
Ref.~\cite{Hartung:2022hoz}. The rotated simple cubical and the
rotated face centred partitionings are new compared to
Ref.~\cite{Hartung:2022hoz}. 

\subsubsection{Genz Points}

We rely on the isomorphism \cref{eq:isomorphy} between $S_3$ and SU$(2)$
to define the set of Genz points for given $m\ge1$ as
\begin{align}
  \begin{split}
    G_m &\coloneqq \left\{\left(s_0\sqrt{\frac{j_0}{m}},s_1\sqrt{\frac{j_1}{m}},s_2\sqrt{\frac{j_2}{m}},s_3\sqrt{\frac{j_3}{m}}\right)\right.\\
    \qquad&\left|\,\sum_{i=0}^3j_i=m,\;\forall i\in \{0,1,2,3\}:\,s_i\in\{\pm1\},\,j_i\in\mathbb{N}\right\}\,.
  \end{split}
\end{align}
This contains all integer partitions $\{ j_0,\dots,j_3\}$ of $m\ge1$
including all permutations and adding all possible sign
combinations.

\subsubsection{Linear Partitioning}

Very similarly, the linear partitioning is defined as the set of
points in $S_3$ based on the same isomorphism
\begin{equation}
  \label{eq:linear_discretisation}
  \begin{split}
    L_m &\coloneqq \left\{\frac{1}{M}\left(s_0j_0,s_1j_1,s_2j_2,s_3j_3\right)\right.\\
    &\left|\,\sum_{i=0}^3j_i=m,\;\forall i\in \{0,1,2,3\}:\,s_i\in\{\pm1\},\,j_i\in\mathbb{N}\right\}\,,
  \end{split}
\end{equation}
with
\begin{equation}
  M \coloneqq \sqrt{\sum_{i=0}^3j_i^2}\,.\label{eq:define_M}
\end{equation}
$M$ takes values $m\geq M \geq \frac{m}{\sqrt{4}}$.

\subsubsection{Volleyball Partitioning}

A variation of $L_m$ is given by the Volleyball partitioning
\begin{align}
  \begin{split}
    V_m &\coloneqq \left\{\frac{1}{M} \left(j_0, j_1, j_2, j_3 \right) \right.\\
    & \left|\,  (j_0, \dots, j_3) \in \left\{ \text{all perm. of } \left( \pm\frac{m}{2}, a_1, \dots, a_3 \right)\right\}, \right. \\
    & \left. \qquad \, a_i \in \{-\frac{m}{2}, -\frac{m}{2} + 1, \dots, \frac{m}{2}\} \vphantom{\frac 1M}\right\}
  \end{split}
\end{align}
with $M$ defined in \cref{eq:define_M}, which takes values $m \le M \le  \sqrt{4} \, m$.
Additionally, the corners of the hypercube, in four dimensions also called $C_8$, form
\begin{align}
  V_0 &\coloneqq \left\{\frac{1}{\sqrt{4}} \left(s_0, \dots, s_3 \right) | \,s_i\in\{\pm1\}\right\}\,,
\end{align}
which is responsible for the name.

\subsubsection{Fibonacci Partitioning}

For a Fibonacci like lattice on $S_3$ we first generate a lattice the
unit cube $[0,1)^3$ defined by
\begin{align*}
  \Lambda_n^{\textrm{Fib}} & = \left\{ t_m \middle| 0 \le m < n, \, \, m \in \mathbb{N} \right\}                                                                                       \\
  t_m & = \begin{pmatrix} t_m^1 \\ t_m^2 \\ t_m^k \end{pmatrix} = \begin{pmatrix}
   \frac{m}{n}        &                      \\
   a_1 \, m \quad     & \mathrm{mod} \quad 1 \\
   a_2 \, m \quad     & \mathrm{mod} \quad 1 \\
  \end{pmatrix}
 \end{align*}
with
\begin{equation}
  \label{eq:fib-irrational-condition}  
  \frac{a_i}{a_j} \notin \mathbb{Q} \quad \textrm{for} \quad i \neq j \textrm{,}
\end{equation}
where $\mathbb{Q}$ denotes the field of rational numbers. In Ref.~\cite{Hartung:2022hoz}
we have chosen $a_1=\sqrt{2}$ and $a_2=\sqrt{3}$. The set of points $\Lambda_n$
can then be mapped to arbitrary manifolds, such as $S_3$. In order to maintain
a uniform density of points, this map needs to volume preserving. In spherical
coordinates, defined by
\begin{equation}\
  z (\psi, \theta, \phi) = \matr{l}{
    \cos \psi\\
    \sin \psi \cos \theta\\
    \sin \psi \sin \theta \cos \phi\\
    \sin \psi \sin \theta \sin \phi}
    \label{eq:spherical-coordinates}
\end{equation}
it can be implemented by the functions
\begin{equation}
  \begin{split}
    \psi(t_m) & =  \Phi_1 (t^1_m) \,,\\
    \theta(t_m) & =  \cos^{-1}\left(1-2 \,t^2_m\right)\\
    \text{and} \qquad \phi(t_m) & = 2 \pi t^3_m \, ,
  \end{split}
  \label{eq:uniform-s3-map}
\end{equation}

where the function $\Phi_1(\psi)$ is defined via its inverse
\begin{equation}
  \Phi_1^{-1} (\psi) = \frac{1}{\pi}  \left( \psi - \frac{1}{2} \sin( 2 \psi) \right)\,.
\end{equation}

\subsubsection{Other Uniform Partitionings}

Finally we can also use the functions defined in \cref{eq:uniform-s3-map} to map
other uniform point sets from the unit cube to the sphere. An obvious choice
would be the simple cubic lattice defined by~\cite{Hahn2001-or}
\begin{equation}
  \Lambda_n^{\textrm{SC}} = \left\{\vec x\in [0,1)^3 \,\middle|\, \vec x =
  d_{\textrm{SC}}(n)\, \textrm{R}\, \vec m
  \,, \vec m \in \mathbb{Z}^3 \right\}
\end{equation}
with a rotation matrix $\textrm{R} \in \textrm{SO}(3)$.
Here $d_{\textrm{SC}}(n)$ denotes the lattice spacing needed
to fit $n$ points into the unit cube 
\[
  d_{\textrm{SC}}(n) = n^{-1 / 3} \,.
\]
The number of sites found inside the cube will be close to $n$.
In general we however still expect a small difference between $n$ and $N\equiv|\Lambda_n^{\textrm{SC}}|$.
While this is no issue
for our application, an exact matching can in principle be achieved by additionally
implementing and tuning a translational offset between the unit cube and the lattice.

The rotation matrix $\textrm{R}$ is used to achieve misalignment
of the lattice planes and the faces of the unit cube. This is required to 
ensure that lattice sites cross the cube boundary individually with varying lattice
spacing. Thus no entire plane of lattice sites is found just in or outside the cube.
For the cubical lattice successive rotations by an angle of $\pi/8$ around $\hat{e}_1$,
$\hat{e}_2$ and $\hat{e}_3$ seem to work well.

To maximise the distance between points we also consider the face centred cubic lattice
given by~\cite{Hahn2001-or}
\begin{equation}
  \begin{split}
  \Lambda_n^{\textrm{FCC}} = \left\{\vec x\in [0,1)^3 \, \middle| \, \vec x =
  \frac{d_{\textrm{FCC}}(n)}{\sqrt{2}} \, \textrm{R} \, \begin{pmatrix} m_1 + m_3 \\ m_2 + m_3 \\ m_1 + m_2 \end{pmatrix}
  , \right. \\ \left. \vec m \in \mathbb{Z}^3 \right\}\,.
\end{split}  
\end{equation}
The lattice spacing for $n$ points is here given by
\begin{equation}
  d_{\textrm{FCC}}(n) = \sqrt[6]{2} \,n^{-1 / 3}
\end{equation}
and maximal as proved in Ref.~\cite{Hales2006}. $R$ is kept from the simple cubical lattice. In the following we will refer to these partitionings as the rotated simple cubical (RSC) and the rotated face centred cubical (RFCC) partitioning respectively.

\subsection{Operator Convergence}
\label{sec:OpConvergence}

The discretised Laplace-Beltrami operator directly enters the
Hamiltonian. Thus, we have to make sure that first of all the spectrum
of the discretised operator converges to the corresponding continuum
spectrum. From the original publication by Kogut and
Susskind~\cite{Kogut:1974ag} the spectrum should be determined by
main angular momentum quantum number $J$ and two independent magnetic quantum
numbers $m_L$ and $m_R$ with $J\ge m_L,m_R\ge 0$.
The two magnetic quantum numbers emerge
because every link connects to two lattice sites with independent
gauge transformations (or left and right gauge transformations).
Therefore, the continuum spectrum is given by
\begin{equation}
  \label{eq:spectrum}
  \mathcal{S}\ =\ \{ J(J+2)\,,\quad J = 0, 1, 2,\ldots\}
\end{equation}
with multiplicity $(J+1)^2$. We note that the eigenvalues are usually
written as $\lambda = j(j+1)$ with $j=0, 1/2, 1,
...$~\cite{Kogut:1974ag}. The form given above is obtained through
rescaling by a factor $4$ and identifying $J=2j$. Then, it matches the spectrum
\cref{eq:spectrum} of the Laplace-Beltrami operator on $S_3$ discussed
above.

In addition to the spectrum we also require that the correct states,
i.e.\ the wave functions, are obtained in the continuum limit. In order to
show this we resort to the concept of convergence in the resolvent
sense. Let $\mathcal{R}$ be the set of resolvents of $-\Delta$. In our case
$\mathcal{R}=\mathbb{C}\setminus\mathcal{S}$, because then with any given $\rho\in
\mathcal{R}$ the inverse of $(\rho-\Delta)$ exists and is bounded from
above. Then, we need to show that
\begin{equation}
  \label{eq:resolvent}
  \lim_{m\to\infty}\| (\rho -\Delta)^{-1} -
  (\rho-\Delta_{\mathcal{D}_m})^{-1} \|\ =\ 0\,,  
\end{equation}
with $\lim_{m\to\infty}\mathcal{D}_m = S_3$. 
Although, the functional analytic convergence analysis of the approximation is shown in \cref{app:gapconvergence}, the analysis in \cref{app:gapconvergence} only implies convergence of the spectrum without any further detail on the rate of convergence. At this point, we will therefore check \cref{eq:resolvent} numerically because the convergence of the spectrum can be estimated against the convergence of resolvents using holomorphic functional calculus for the spectral projectors~\cite{Hartung-PhD}.

To do so, we first introduce the eigenfunctions $Y_{J, l_1, l_2}$ of $-\Delta$ in the continuum. They are given by the spherical harmonics in four dimensions~\cite{erdelyi1953bateman}.

Note that $J$ is the same as above and the eigenvalues of the continuum $\Delta$ are independently
of $l_1$ and $l_2$ given by $\lambda = J(J + 2)$. The $l_{1,2}$ are not identical with the magnetic quantum numbers $m_{L,R}$, however. Instead, the obey the condition $J \geq l_1 \geq |l_2|$.
It can readily be checked that this condition leads to the same
multiplicity $(J+1)^2$ as above. 

The spherical harmonics can be expressed in terms of the spherical coordinates introduced in \cref{eq:spherical-coordinates} as
\begin{equation}
  Y_{J, l_1, l_2} =  \frac{\mathrm{e}^{i l_2 \varphi}}{\sqrt{2}} \,  {}_2 \overline{P}^{|l_2|}_{l_1} (\theta) \,  {}_3\overline{P}^{l_1}_{J} (\psi) \,,
  \label{eq:spherical-harmonics}
\end{equation}
where
\begin{equation}
  \begin{split}
    {}_n \overline{P}^{l}_L (\zeta) = \sqrt{\frac{2L+n-1}{2} \frac{(L+l+n-2)!}{(L-l)!}} & \\ \times\frac{1}{(\sin \zeta)^{(n-2)/2}} \, P^{-\left(l+\frac{n-2}{2}\right)}_{L+\frac{n-2}{2}} & (\cos \zeta)\,.
  \end{split}
\end{equation}
$P^{-\mu}_{\nu} (x)$ here denote the Legendre functions of first kind and are given in terms of the hyper-geometric function ${}_2 F_1$ as
\begin{equation}
  \begin{split}
    P^{-\mu}_{\nu} (x) = \frac{1}{\Gamma(1 + \mu)} & \left( \frac{1-x}{1+x} \right)^{\mu/2} \\
      {}_2F_1 & \left(-\nu, \nu +1; 1+\mu; \frac{1-x}{2}\right) .
  \end{split}
\end{equation}
Similarly to their famous counterparts on $S_2$, these form an orthonormal basis of the square integrable functions on $S_3$. Thus we can express the discretised operator $\Delta_{\mathcal{D}_m}$ in this basis by evaluating
\begin{equation}
  \begin{split}
    \langle Y_{J,l_1,l_2}, \Delta_{\mathcal{D}_m} Y_{J',l_1',l_2'} \rangle &= \\ 
    \int_{S_3} \mathrm{d}V \, & Y_{J,l_1,l_2} \, \Delta_{\mathcal{D}_m} Y_{J',l_1',l_2'}\,.
  \end{split}
\end{equation}
Similarly to \cref{eq:laplacian-rhs-M}, we can approximate this integral
numerically by evaluating the spherical harmonics at the vertices of
$\mathcal{D}_m$ and weighting them with the corresponding barycentric cell volume $v(j)$.

If we then implement an upper limit for $J$ by imposing $J < J^{\max}$ we can finally evaluate and check \cref{eq:resolvent}. Such a truncation is acceptable, as bigger $J$ correspond to bigger eigenvalues $\lambda = J(J + 2)$ leading to decreasing contributions to the inverse operator $(\rho - \Delta)^{-1}$.

As the operator norm $\| \cdot \|$ of an operator $O$ we choose 
\begin{equation}
  \|O\| = \sqrt{\lambda_{\max}}
\end{equation}
where $\lambda_{\max}$ denotes the biggest eigenvalue of $O^{\dagger} O$.
 
\section{Numerical Experiments}

In the following, we benchmark the performance of several representative partitionings as applied to different observables. A complete list including all the partitionings we considered for this work can be found in \cref{sec:all_partitionings}.

\begin{figure}
  \includegraphics[width=0.95\linewidth]{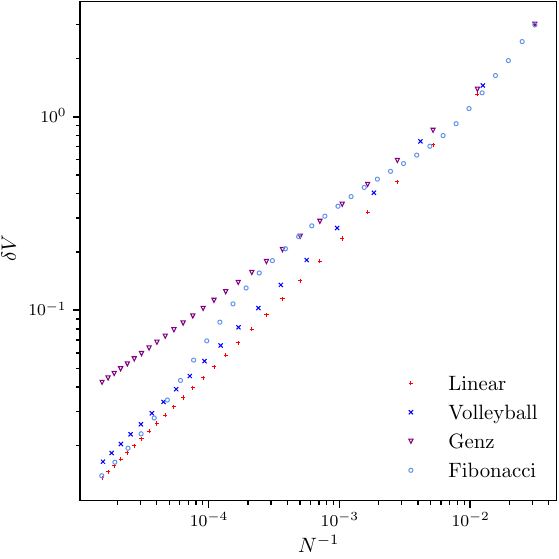}
  \caption{The volume of $S_3$ minus the sum of the barycentric cell volumes
    as a function of $1/N$ in a double log plot. Here, $N$ is the number of 
    points in the partitioning. We compare Fibonacci, Genz, Linear and
    Volleyball partitionings as indicated in the legend.}
  \label{fig:volConvergenceV1}
\end{figure}

\subsection{Volume Convergence}

In \cref{sec:triangulation} we have shown how to construct the
Laplace-Beltrami operator based on a triangulation procedure in $S_3$
based on the partitioning. For this we have used the barycentric
cell volumes $v(j)$ according to \cref{eq:volbary}. The sum of all
$v(j)$ approximates the volume of $S_3$, and by increasing the
number of points in the partitioning we expect convergence towards
Vol$(S_3)$.

The speed of convergence will certainly depend on the actual
partitioning, and we expect this in turn to influence the
approximation of the Laplace-Beltrami operator. This is in particular
so because we have used Greens theorem neglecting the fact that we
only approximate $S_3$.

In \cref{fig:volConvergenceV1} we plot
\[
\delta V\ =\ \mathrm{Vol}(S_3) - \sum_j v(j)
\]
as a function of $1/N$ for different partitionings in a double logarithmic
plot, where we recall that $N$ is the number of elements in the
respective partitioning. For the Genz, the Linear and the Volleyball partitionings we
observe that the missing volume $\delta V$ is proportional to $N^{-y}$
with $y\approx 2/3$ for Linear and Volleyball partitionings and
$y\approx 1/2$ for the Genz points.

Since the mean distance between two points is roughly $N^{-1/3}$, we
conclude that convergence towards $S_3$ appears to be quadratic in
the mean distance for the Linear and the Volleyball
partitioning. For the Genz points on the other hand convergence is
proportional to the mean distance to the power of $3/2$, and thus
significantly slower. We recall from~\cite{Hartung:2022hoz} that the
distance between neighbouring points is $1/m$ for the Linear
partitioning $L_m$ mostly independent of the direction, whereas it is
between $1/m$ and $\sqrt{2/m}$ for the Genz partitioning $G_m$
depending on the direction and the point itself. In particular, the
weights of the different points (derived from the associated volume)
differ by up to a factor $m^{3/2}$ for 
$G_m$, while the corresponding weight factor for $L_m$ is independent of
$m$. Thus, Genz points are significantly less isotropically
distributed on $S_3$, which is responsible for the slower
convergence. 

The Fibonacci partitioning with $a_1=\sqrt{2}$ and $a_2=\sqrt{3}$
fixed for all $N$ shows an irregular convergence behaviour: for small
$N$ it seems to be similar to the Linear partitioning, then it appears
to converge like the Genz points for intermediate $N$-values, but gets
in line again with the Linear partitioning for large $N$. Similar behaviour
is also observed for other choices for $a_1$ and $a_2$, only the location of
the bump changes.

The reason for this is the following: for a poor choice of the coefficients
$a_1$ and $a_2$ the modulo operation in $t_m^2$ and $t_m^3$ can lead to
an almost periodic behaviour, such that $t_m^{2/3} \approx t_{m+p}^{2/3}$. If
the period $p$ is small, this means that the points $t_m$ and $t_{m+p}$ end
up close together in the cube and on $S_3$. Having multiple lattice sites
in almost the same spot will do little to improve the quality of our Riemann
sum and thus leads to the slowing down in convergence. The return to the 
convergence rate of the linear and Volleyball partitionings towards larger 
$N$ is observed, because \cref{eq:fib-irrational-condition} ensures that 
$t_m^{2/3} \neq t_{m+p}^{2/3}$. Their non-zero difference then becomes more 
and more significant with finer lattice spacing, and thus eventually removes 
the periodicity in the coordinates.

\begin{figure}
  \includegraphics[width=0.95\linewidth]{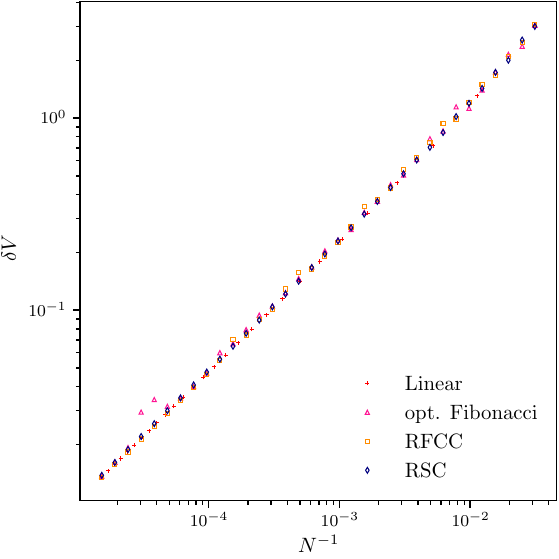}
  \caption{Like \cref{fig:volConvergenceV1}, but comparing the
    original Fibonacci partitioning with the optimised one and the RFCC
    and RSC partitionings. }
  \label{fig:volConvergenceV2}
\end{figure}

In order to overcome this irregular convergence pattern of the
original Fibonacci partitioning, we choose $a_1$ and $a_2$
separately for each $N$. The distance of a potential close neighbour with period $p$ can be calculated as
\begin{equation}
  \begin{split}
    \delta_N(p,a_1,a_2) &= \left[ (p/N)^2 + ( p a_1 \mod 
    1)^2 \right. \\ 
    & \qquad \qquad \qquad \left. + ( p a_2 \mod 1)^2 \right]^{1/2} \, .
  \end{split}
\end{equation}
An upper bound for $p$ is given by
\[
  \frac{p}{N} < d_{\textrm{FCC}}(N)  
\]
as the face centred cubical packing maximises the distance between
lattice points. Thus we can predict the distance to the closest neighbour
for two coefficients $a_1$ and $a_2$ by evaluating 
\begin{equation}
  d_{\textrm{Fib}} (N, a_1, a_2) = \min \left( \left\{ \delta_N (p,a_1,a_2) \, \middle| \, p \in \mathbb{N} \setminus \{0\} \right\}  \right)\,.
\end{equation}
With this we can simply iterate over a suitable set of coefficients
$a_{1/2}$ until the desired minimum distance is reached. In our
implementation we iterated over the square roots of the prime numbers
until
\begin{equation}
  \frac{d_{\textrm{Fib}}(n, a_1, a_2)}{d_{\textrm{FCC}}(n)} \geq 0.95
\end{equation}
was fulfilled.

The volume convergence of the original and the such optimised
Fibonacci partitionings are compared in \cref{fig:volConvergenceV2}. 
We observe that the irregularities are mostly gone in the optimised
version. However, since $a_1$ and $a_2$ parameters are optimised for
each $N$, one can expect a uniform convergence rate, but the amplitude
might still depend on $N$, which is what we see in the form of outliers. In the following we,
therefore, use the Fibonacci partitioning with optimised values for
$a_1$ and $a_2$ only. 

This remaining irregularity can be cured by using the RFCC or the RSC
partitionings, as also shown in \cref{fig:volConvergenceV2}. The
smallest overall deviations from the volume of $S_3$ are actually
observed for the RFCC partitioning, but the RSC does not perform
significantly worse.

\begin{figure}
  \centering
  \includegraphics[width=.95\linewidth]{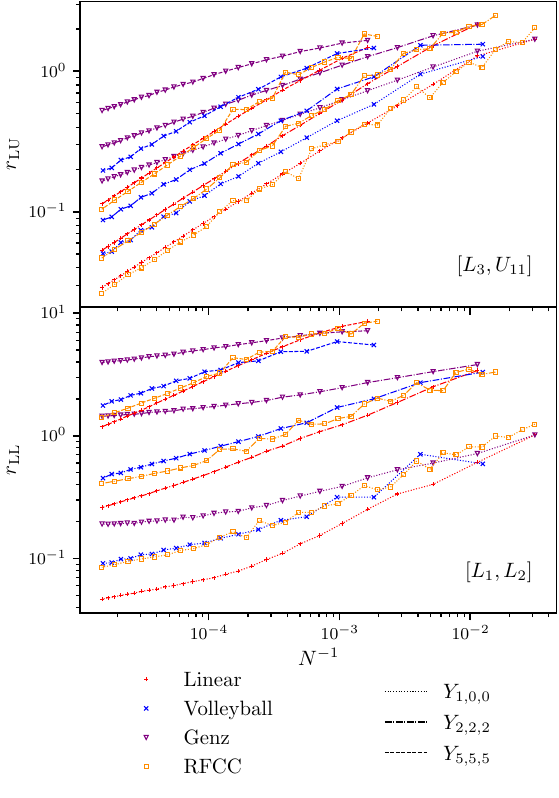}
  \caption{We show
    $r_\mathrm{LU}$ (upper panel) and $r_\mathrm{LL}$ (lower panel) as
  a function of $1/N$ for three different spherical harmonics in a
  double log plot. We compare Linear, Volleyball, Genz and RFCC
  partitionings.} 
  \label{fig:commConvergence}
\end{figure}

\subsection{Commutation Relations}

With the above definitions of $\hat L$ and $\hat R$ it is ensured that
if applied to a constant vector one obtains zero. Much like in the one
dimensional case of a finite difference operator, we expect $\hat L$
and $\hat R$ to work best if applied to slowly varying vectors in the
algebra. Thus we choose some of the lower lying spherical harmonics
$Y_{J , l_1 ,l_2}$ defined in \cref{eq:spherical-harmonics} as test
functions. For each harmonic we can compute the corresponding error
vector
\begin{equation}
  \label{eq:zLU}
  w\ =\ \left([L_a, U_{jl}] -(t_a)_{ji} U_{il}\right) \cdot Y_{J , l_1 ,l_2} 
\end{equation}
and then the mean deviation weighted by barycentric cell volume $v(i)$ as
\begin{equation}
  r_\mathrm{LU}\ =\ \sum_i v(i) \, |w_i|\,.
\end{equation}
Likewise, we define
\begin{equation}
  \label{eq:zLL}
  u\ =\ \left([L_a, L_b]\ +\ 2
  \mathrm{i}\,\epsilon_{abc}\,L_c\right)\cdot  Y_{J , l_1 ,l_2}  \,,
\end{equation}
with the appropriate structure constant $f_{abc}$ for SU$(2)$, and
\begin{equation}
  r_\mathrm{LL}\ =\ \sum_i v(i) \, |u_i|\,.
\end{equation}
In \cref{fig:commConvergence} we plot $r_\mathrm{LU}$ and
$r_\mathrm{LL}$ for exemplary combinations of indices $j$ and $l$ as a
function of $1/N$. Results are shown for the different
partitionings of SU$(2)$ and different spherical harmonics. 

We observe that $r_\mathrm{LU}$ approaches zero with increasing
$N$. The convergence rate appears to be depending on the partitioning
used: while Genz points seem to converge the slowest, the Linear and
RFCC partitionings work best in this respect. As expected the
deviations increase when using faster oscillating
harmonics.

For $r_\mathrm{LL}$, and thus the deviations from the expected
behaviour in the commutator of $L$ with itself, the picture is less
clear. We again see a decrease in $r_\mathrm{LL}$ with increasing $N$,
however, the convergence rate appears slower and the scaling region
might set in only at larger $N$ compared to $r_\mathrm{LU}$.

Interestingly, the Linear partitioning shows the fastest
convergence for $r_\mathrm{LL}$ compared to all the other
partitionings investigated here.

\begin{figure}
  \centering
  \includegraphics[width=\linewidth]{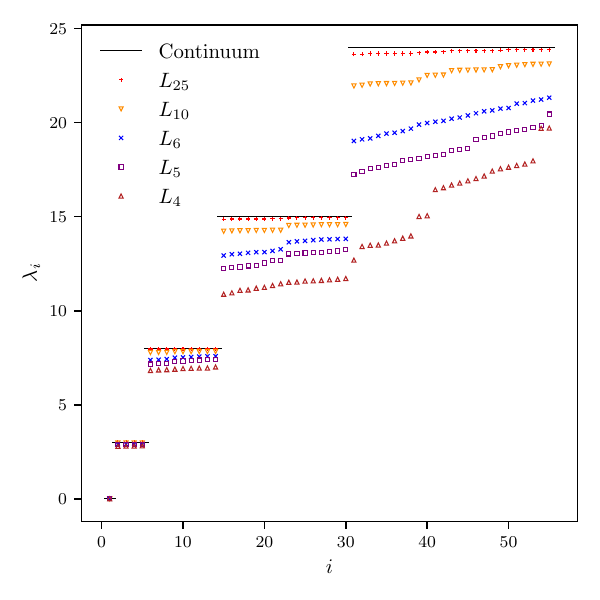}
  \caption{We show the lowest 60 eigenvalues of the discretised
    Laplace-Beltrami operator for the Linear partitionings $L_m$ with
    $m=4,5,6,10$ and $m=25$. The solid lines corresponds to the expected
    continuum values from \cref{eq:spectrum}.}
  \label{fig:specOverview}
\end{figure}

\subsection{Spectrum in the Free Theory}

We compute the spectrum of the discretised Laplace-Beltrami operator
$-\Delta_{\mathcal{D}_m}$ for different partitionings $\mathcal{D}$ and
values of $m$ numerically. We note that the lowest eigenvalue
$\lambda_1=0$ per construction, and therefore we are going to mostly
exclude $\lambda_1$ from the following discussion.

For the linear partitioning we show the
60 lowest eigenvalues $\lambda_i$ in \cref{fig:specOverview}. While
the different point styles distinguish values of $m$ from $4$ to $25$,
the line indicates the continuum spectrum from \cref{eq:spectrum}. It is
clearly visible that with increasing $m$ the spectrum of
$-\Delta_{L_m}$ converges towards the continuum spectrum with the
correct multiplicity.

\begin{figure}
  \centering
  \includegraphics[width=\linewidth]{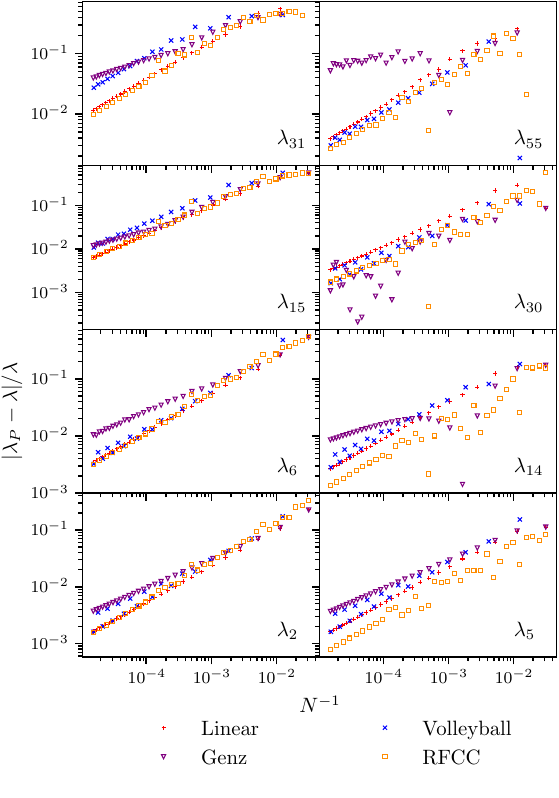}
  \caption{We plot $|\lambda_\mathrm{P}- \lambda|/\lambda$ as a
    function of $1/N$ for the Volleyball, the Linear, the RFCC
    and the Genz partitionings for the eigenvalues $\lambda_i$ with
    $i=2, 5, 6, 14, 15, 30, 31, 55$.}
  \label{fig:specConvergenceLog}
\end{figure}

In \cref{fig:specConvergenceLog} we show the
convergence for different eigenvalues $\lambda_\mathrm{P}$ of the discretised
Laplace-Beltrami operator separately. We plot the relative
deviation from the expected continuum value $\lambda$
\[
\frac{|\lambda_\mathrm{P} -\lambda|}{\lambda}
\]
as a function of $1/N$ for different eigenvalues $\lambda_k$ with
$k=2, 5, 6, 14, 15, 30, 31, 55$. Note that we order the eigenvalues
such that $\lambda_i \leq \lambda_{i+1}$. Our choices for $i$,
therefore, correspond to the first and the last eigenvalue in a
multiplet. We observe convergence towards the expected continuum
value for all the eigenvalues we investigated and all the partitionings.

The most regular and smooth convergence pattern is observed for the
linear partitioning. The Volleyball partitioning behaves similarly,
but with some dependence on the actual value of $N$.
The convergence rate of the RFCC partitioning falls in
line with the one of the Linear partitioning, but with somewhat
smaller amplitude.

The Genz partitioning seems to be also converging, however, with a
visibly slower convergence rate than the other partitionings towards
large $N$-values. In particular for the last eigenvalues of a given
multiplet we also observe sign changes in
$\lambda_\mathrm{P}-\lambda$.

Empirically, the convergence rates appear to be independent of the
index $i$ and very similar to the rates of convergence of the volume.

\subsection{Operator Convergence}

\begin{figure}[t]
  \includegraphics[width=0.95\linewidth]{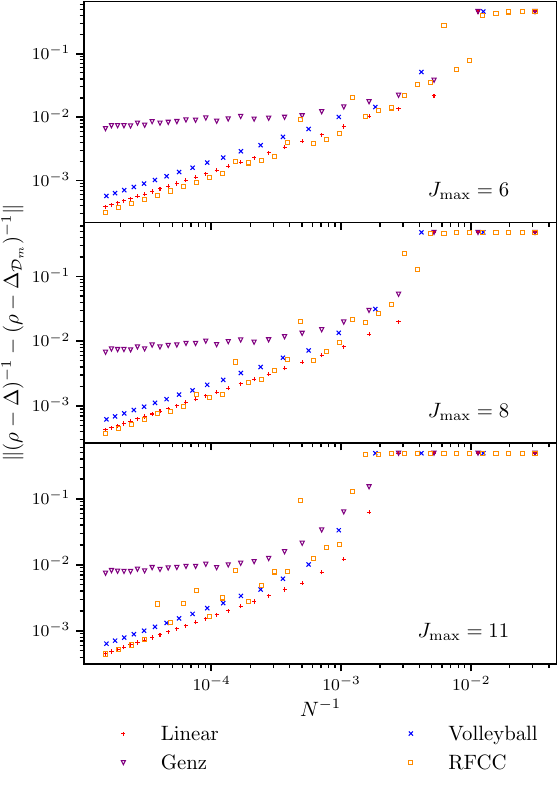}
  \caption{We plot 
    $\|(\rho-\Delta)^{-1}-(\rho-\Delta_{\mathcal{D}_m})^{-1}\|$ as a
    function of $1/N$ evaluated on different sets of eigenfunctions
    corresponding to $J_\mathrm{max}=6$, $J_\mathrm{max}=8$ and
    $J_\mathrm{max}=11$ with $\rho=-2$.}
  \label{fig:gapConvergence}
\end{figure}

Finally, we discuss the operator convergence by checking
\cref{eq:resolvent} numerically.
For this we pick
a value from the resolvent set of $\rho=-2$. Other $\rho$-values lead
to similar and qualitatively equivalent results.

In \cref{fig:gapConvergence} we plot
\begin{equation}
  \label{eq:gap}
  \|(\rho-\Delta)^{-1}-(\rho-\Delta_{\mathcal{D}_m(N)})^{-1}\|  
\end{equation}
as a function of $1/N$ for $\rho=-2$. 
As discussed in
\cref{sec:OpConvergence}, we evaluate \cref{eq:gap} using a finite subset of the spherical
harmonics $Y_{J, l_1, l_2}$ by imposing $J < J_\mathrm{max}$.
The three different $J_\mathrm{max}$-values are
$J_\mathrm{max}=6$, $J_\mathrm{max}=8$ and $J_\mathrm{max}=11$. 

Concentrating on the uppermost panel for $J_\mathrm{max}=6$ to start
with, we observe gap convergence for all partitionings shown.
The convergence rate is proportional to
$N^{-z}$ with $z\approx 0.7$ for all partitionings apart from
Genz. For the Genz partitioning the convergence is 
very slow with $z\approx 0.1$. This picture remains the
same for large $N$-values for the two other $J_\mathrm{max}$-values.

For $J_\mathrm{max}=8$ and $J_\mathrm{max}=11$ another feature becomes
visible: for $N\to0$ the gap plateaus at a value of $1/|\rho|$. At the
corresponding values of $N$ there are not sufficiently many points in
the partitioning to resolve all the $Y_{J, l_1, l_2}$.

\section{Discussion}

There are a few points that deserve discussion. First, the
results from the previous section indicate that the points in a
partitioning should be as uniformly distributed as possible: the main
difference between the Genz $G_m$ and the Linear $L_m$ partitionings
is that in the Genz set points are denser around the
poles. This leads to larger deviations from the mean distance for
$G_m$ compared to $L_m$ at fixed $m$, which in turn means slower
convergence to the volume of $S_3$. This conclusion is strongly
supported by the comparison of original Fibonacci and optimised
Fibonacci partitioning, where the difference in convergence can be
clearly traced back to the non-uniformity in the original Fibonacci
version for certain values of $N$. For additional results supporting
this conclusion we refer to \cref{sec:all_partitionings}.

Since we define the Laplace-Beltrami operator via an integral relation
\cref{eq:Lsq}, the aforementioned effects from non-uniformity are
expected to also influence the spectrum of the discretised operator as
well as the gap convergence. This explains the slower
convergence rates observed for the Genz partitioning for most of the
investigated quantities.

In addition, we observe plateaus in the gap in
\cref{fig:gapConvergence} for small values of $N$. The extent of the
plateaus depends on $J_\mathrm{max}$. They originate from a
sufficiently large $N$ being required to resolve all the states up to
a given $J_\mathrm{max}$.
More specifically, $J$ corresponds to an (angular) momentum. The
highest momentum $J_\text{max}$ that can be resolved on a lattice, is
dictated by the inverse lattice spacing. 
Put differently, the lattice discretisation acts as an ultraviolet cutoff.
Since the lattice spacing is proportional to $N^{-1/3}$, the plateaus
are expected for $N\lesssim J_\text{max}^3$. This is well compatible
with \cref{fig:gapConvergence}. 

\section{Conclusion and Outlook}

In this paper we have shown how to discretise the electric part in the
Hamiltonian \cref{eq:hamiltonian} for the gauge group SU$(2)$, when
the basis is chosen such that the gauge field operators $\hat U$ are
diagonal. It turns out that the canonical momentum operators $\hat L$
and $\hat R$ can be constructed by discretising the corresponding Lie
derivatives based on a triangulated partitioning of SU$(2)$. Mostly
independently on the choice of the partitioning the such constructed
operators fulfil the fundamental commutation relations up to
discretisation effects. However, when it comes to reproducing the free
spectrum of the theory, it is not sufficient to insert these $\hat L$
and $\hat R$ squared into the Hamiltonian.

It is rather necessary to construct a discrete version of the electric part
in the Hamiltonian directly by realising that it corresponds to the
Laplace-Beltrami operator on $S_3$. This operator can be discretised
by means known from finite element methods.

Then, our results show that with sufficiently uniform
partitionings the low lying eigenvalues of the discretised
Laplace-Beltrami operator converge towards their continuum
counterparts. The larger the number of points $N$ in the partitioning, 
the more eigenvalues can be resolved. Likewise, the continuum wave
functions are reproduced with $N\to\infty$. Thus, we conclude that the
discretised free Hamiltonian reproduces the free theory up to
discretisation effects. The size of these artefacts can be reduced
arbitrarily by increasing $N$.

In the future we will investigate the SU$(2)$ discretisation proposed
in this paper beyond the free theory. Moreover, the implementation of
Gauss' law and the consequences of the breaking of the fundamental
commutation relations (albeit only by discretisation effects) will be
important to understand.

\begin{acknowledgments}

  We thank A.~Crippa, G.~Clemente, J.~Haase and M.~A.~Schweitzer for helpful discussions.
  This work is supported by the Deutsche
  Forschungsgemeinschaft (DFG, German Research Foundation) and the NSFC through the funds provided to the Sino-German
  Collaborative Research Center CRC 110 “Symmetries
  and the Emergence of Structure in QCD” (DFG Project-ID 196253076 -
  TRR 110, NSFC Grant No.~12070131001)
  as well as the STFC Consolidated Grant ST/T000988/1.
  This work is supported with funds from the Ministry of Science,
  Research and Culture of the State of Brandenburg within the Centre
  for Quantum Technologies and Applications (CQTA). This work is
  funded by the European Union’s Horizon Europe Framework Programme
  (HORIZON) under the ERA Chair scheme with grant agreement No. 101087126.
  \flushright{\includegraphics[width=0.08\textwidth]{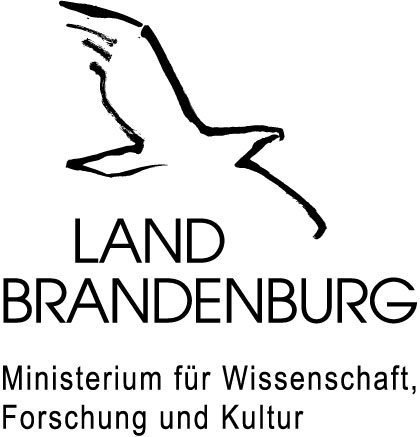}}
\end{acknowledgments}

\appendix{
  \newcommand{\eps}{\ensuremath{\varepsilon}}
\newcommand{\vphi}{\ensuremath{\phi}}
\renewcommand{\phi}{\ensuremath{\varphi}}
\newcommand{\subdot}{\d}
\renewcommand{\d}{\partial}
\newcommand{\ex}{\exists}
\newcommand{\fa}{\forall}
\newcommand{\lnorm}{\left\lVert}
\newcommand{\rnorm}{\right\lVert}
\newcommand{\lbetr}{\left\lvert}
\newcommand{\rbetr}{\right\lvert}
\newcommand{\norm}[1]{\lnorm {#1}\rnorm}
\newcommand{\pr}{\operatorname{pr}}
\newcommand{\dist}{\mathrm{dist}}

\section{Functional analytic point of view on convergence}
\label{app:gapconvergence}
\subsection{Domain of the discretized derivative and Laplacian}
We are generally looking for a discretisation of the Laplace-Beltrami in $L_2(\mathrm{SU}(2))$. However the discretised derivative $\nabla_P$ -- as constructed following \cref{subsec:construction-gradient} using a triangulated partition $P$ of $\mathrm{SU}(2)$ -- requires point-evaluations on the vertices of the partitioning $P$. Hence, we can only a priori define $\nabla_P$ on functions in $C(\mathrm{SU}(2))$, and then need to consider the $L_2(\mathrm{SU}(2))$ extension of the thus defined $\nabla_P|_{C(\mathrm{SU}(2))}$.

Using the Sobolev Embedding for compact manifolds without boundary\footnote{Let $M$ be a compact manifold without boundary, $k,r\in\mathbb{N}_0$, $\alpha\in[0,1)$, $p\ge1$, and $\frac{k-r-\alpha}{\dim M}\ge\frac{1}{p}$. Then, $W_p^k(M)\subseteq C^{r,\alpha}(M)$.} we obtain the (tightest) embeddings
\begin{align}
  \fa k\ge2:\ W_2^k(\mathrm{SU}(2))\subseteq C^{k-2,\frac12}(\mathrm{SU}(2)).
\end{align}
In particular, this implies that we can define $\nabla_P|_{W_2^2(\mathrm{SU}(2))}$ using the point-evaluation formula. A priori this means that $\nabla_P|_{W_2^2(\mathrm{SU}(2))}$ maps $W_2^2(\mathrm{SU}(2))$ into $L_2(\mathrm{SU}(2))$ instead of the expected $W_2^1(\mathrm{SU}(2))$. It is now important to note that, in local coordinates, difference quotients $D_i^h$ with step size $h$ in direction $i$ satisfy $\norm{D_i^hu}_{L_p(\Omega')}\le \norm{D_iu}_{L_p(\Omega)}$ for $\Omega'\Subset\Omega$ provided $h$ is sufficiently small. Lifting this to $\mathrm{SU}(2)$ means that $\nabla_P|_{W_2^2(\mathrm{SU}(2))}$ is relatively bounded by $\nabla|_{W_2^2(\mathrm{SU}(2))}$, i.e., $\nabla_P|_{W_2^2(\mathrm{SU}(2))}$ in fact maps $W_2^2(\mathrm{SU}(2))$ into $W_2^1(\mathrm{SU}(2))$ as expected and can be uniquely extended to the operator $\nabla_P:\ W_2^1(\mathrm{SU}(2))\to L_2(\mathrm{SU}(2))$.

Finally, this implies that the discretised Laplace $\Delta_P$ defined as the operator associated with the symmetric form $\tau(x,y)=\langle \nabla_Px,\nabla_Py\rangle$ is a well-defined map $\Delta_P:\ W_2^2(\mathrm{SU}(2))\to L_2(\mathrm{SU}(2))$.

\subsection{Pointwise Convergence}

Further to the discretised gradient $D^h$ defined via difference quotients in local coordinates being bounded by the gradient $\nabla$, $D^hf\to\nabla f$ holds in $W_p^{k-1}(\Omega)$ for any $f\in W_p^k(\Omega)$. Hence, again lifting this result to $\mathrm{SU}(2)$, we obtain pointwise convergence $\nabla_P\to\nabla$ using the net of discretisations with directed set of partitionings $P'\preceq P$ if and only if every vertex of $P'$ is also a vertex of $P$. This also implies that the discretised Laplace $\Delta_P=\nabla_P^*\nabla_P:\ W_2^2(\mathrm{SU}(2))\to L_2(\mathrm{SU}(2))$ converges pointwise to the Laplace $\Delta=\nabla^*\nabla:\ W_2^2(\mathrm{SU}(2))\to L_2(\mathrm{SU}(2))$.

Unfortunately, this pointwise convergence is not sufficient for convergence of the spectrum. To obtain the convergence of the spectrum, we need a notion called gap convergence which is equivalent to norm convergence for bounded operators and convergence in norm resolvent sense for closed operators with non-empty resolvent set \cite{Kato}. Using \cref{eq:resolvent}, we have tested the convergence in norm resolvent sense numerically. 

\subsection{Finite order gap convergence}
Let us consider an orthonormal basis $(\psi_j)_{j\in\mathbb{N}}$ of $L_2(\mathrm{SU}(2))$ with $\fa j\in\mathbb{N}:\ \psi_j\in W_2^2(\mathrm{SU}(2))$. Restricting the space $L_2(\mathrm{SU}(2))$ and $W_2^2(\mathrm{SU}(2))$ to the linear span $L_2(\mathrm{SU}(2))|_n$ (and $W_2^2(\mathrm{SU}(2))|_n$ with the respective topology) of the first $n$ basis vectors induces restricted operators\footnote{We will suppress the projections $\pr_{L_2(\mathrm{SU}(2))|_n}$ for brevity, i.e., we will simply write $\Delta_P|_n$ and $\Delta|_n$ instead of $\pr_{L_2(\mathrm{SU}(2))|_n}\Delta_P|_n$ and $\pr_{L_2(\mathrm{SU}(2))|_n}\Delta|_n$. This has no impact on the following estimates since $\norm{\pr_{L_2(\mathrm{SU}(2))|_n}x}\le\norm{x}$.} $\Delta_P|_n$ and $\Delta|_n$. For these, we obtain
\begin{align}
  \begin{aligned}
    &\norm{\Delta_P|_n-\Delta|_n}_{L(W_2^2(\mathrm{SU}(2))|_n,L_2(\mathrm{SU}(2))|_n)}\\
    =&\sup_{\norm{\phi}_{W_2^2(\mathrm{SU}(2))|_n}=1}\norm{\Delta_P|_n\phi-\Delta|_n\phi}_{L_2(\mathrm{SU}(2))|_n}\\
    =&\sup_{\norm{\sum_{j=1}^n\alpha_j\psi_j}_{W_2^2|_n}=1}\norm{(\Delta_P|_n-\Delta|_n)\sum_{j=1}^n\alpha_j\psi_j}_{L_2|_n}\\
    \le&\sup_{\norm{\sum_{j=1}^n\alpha_j\psi_j}_{W_2^2|_n}}\sum_{j=1}^n\abs{\alpha_j}\norm{(\Delta_P|_n-\Delta|_n)\psi_j}_{L_2|_n}.
  \end{aligned}
\end{align}
Since $\norm\cdot_{W_2^2(\mathrm{SU}(2))}\ge\norm\cdot_{L_2(\mathrm{SU}(2))}$ we observe for $\phi=\sum_{j=1}^n\alpha_j\psi_j\in\d B_{W_2^2(\mathrm{SU}(2))|_n}$
\begin{align}
  \begin{aligned}
    \sum_{j=1}^n\abs{\alpha_j}=&\norm\alpha_{\ell_1(n)}\le\sqrt{n}\norm\alpha_{\ell_2(n)}=\sqrt{n}\norm{\phi}_{L_2}\\
    \le&\sqrt{n}\norm{\phi}_{W_2^2}=\sqrt{n}
  \end{aligned}
\end{align}
and thus
\begin{align}
  \begin{aligned}
    \norm{\Delta_P|_n-\Delta|_n}
    \le&\sqrt{n}\underbrace{\max_{j\le n}\norm{(\Delta_P|_n-\Delta|_n)\psi_j}_{L_2|_n}}_{\to0\ (P\nearrow)}.
  \end{aligned}
\end{align}
Since norm convergence and gap convergence are equivalent for bounded operators, we can conclude that the restricted discretised Laplace converges in gap to the restricted continuum Laplace.

\subsection{$L_2$ with decay}
Unfortunately, this gap convergence seems not to extend to all of $L_2(\mathrm{SU}(2))$. Instead, we consider a conic submanifold of ``$L_2$-functions with decay''. A function $\phi\in L_2(\mathrm{SU}(2))$ is an element of $L_{2,0(\beta,\delta)}(\mathrm{SU}(2))$ if and only if for a given orthonormal basis $\beta=(\psi_j)_{j\in\mathbb{N}}$ of $L_2(\mathrm{SU}(2))$ and a decay function $\delta:\ \mathbb{N}\to\mathbb{R}_{>0}$ with $\delta(n)\to0\ (n\to\infty)$
\begin{align}
  \fa n\in\mathbb{N}:\ \norm{\phi-\sum_{j=1}^n\langle\psi_j,\phi\rangle\psi_j}_{L_2(\mathrm{SU}(2))}\le\delta(n)\norm{\phi}
\end{align}
holds. In other words, we restrict the space of functions to those whose Fourier modes of order $n$ and higher contribute no more than $\delta(n-1)$ in norm. In this sense, this is similar to a UV cutoff although not quite as strong as still arbitrarily large Fourier modes are allowed, they just cannot have arbitrarily large Fourier coefficients.

For given $n$, we will use the notation 
\begin{align}
  \phi^\downarrow=&\sum_{j=1}^n\langle\psi_j,\phi\rangle\psi_j\text{ and }\phi^\uparrow=\sum_{j=n+1}^\infty\langle\psi_j,\phi\rangle\psi_j.
\end{align}

We note that this $L_{2,0(\beta,\delta)}(\mathrm{SU}(2))$ is a closed conic submanifold of $L_2(\mathrm{SU}(2))$ since for $\phi\in L_{2,0(\beta,\delta)}(\mathrm{SU}(2))$ and $r>0$ we obtain $r\phi\in L_{2,0(\beta,\delta)}(\mathrm{SU}(2))$  (conic), and for $\phi_k\in L_{2,0(\beta,\delta)}(\mathrm{SU}(2))$ and $\phi_k\to\phi$ in $L_2(\mathrm{SU}(2))$ we have
\begin{align}\label{eq:closedL2}
  \begin{aligned}
    &\norm{\phi-\sum_{j=1}^n\langle\psi_j,\phi\rangle\psi_j}\\
    \le&\norm{\phi-\phi_k}+\norm{\phi_k-\sum_{j=1}^n\langle\psi_j,\phi_k\rangle\psi_j}\\
    &+\norm{\sum_{j=1}^n\langle\psi_j,\phi_k-\phi\rangle\psi_j}\\
    \le&2\underbrace{\norm{\phi-\phi_k}}_{\to0}+\underbrace{\norm{\phi_k-\sum_{j=1}^n\langle\psi_j,\phi_k\rangle\psi_j}}_{\le \delta(n)\norm{\phi_k}}\\
    \le& \delta(n)\norm{\phi_k}+o(1)\\
    \to& \delta(n)\norm{\phi},
  \end{aligned}
\end{align}
i.e., $L_{2,0(\beta,\delta)}(\mathrm{SU}(2))$ is closed.

\subsection{Gap convergence with decay}

Unfortunately, $L_{2,0(\beta,\delta)}(\mathrm{SU}(2))$ is not a linear subspace since sums of its elements is in general not elements of $L_{2,0(\beta,\delta)}(\mathrm{SU}(2))$. However, it has the induced topology from $L_2(\mathrm{SU}(2))$ and thus the $L(L_2(\mathrm{SU}(2)))$ and $L(W_2^2(\mathrm{SU}(2)),L_2(\mathrm{SU}(2)))$ norms computed by restricting the unit sphere of $L_2(\mathrm{SU}(2))$ and $W_2^2(\mathrm{SU}(2))$ respectively are consistent with the operator norms in $L(L_{2,0(\beta,\delta)}(\mathrm{SU}(2)))$ and $L(W^2_{2,0(\beta,\delta)}(\mathrm{SU}(2)),L_{2,0(\beta,\delta)}(\mathrm{SU}(2)))$. Here, we denote the restriction to $W^2_{2,0(\beta,\delta)}(\mathrm{SU}(2))$ as the set of $L_2(\mathrm{SU}(2))$ functions $\phi$ whose derivatives up to second order are in $L_{2,0(\beta,\delta)}(\mathrm{SU}(2))$ including the condition that derivatives up to second order of $\phi^\uparrow$ are bounded by $\delta$. This is still a closed conic submanifold of $L_2(\mathrm{SU}(2))$ and of $W_2^2(\mathrm{SU}(2))$ using the same argument as in \cref{eq:closedL2} changing only the norm. It should be noted that this requires $\beta$ to be an orthonormal basis that is contained in $W_2^2(\mathrm{SU}(2))$. In our case, this basis given by the spherical harmonics which are in $W_2^\infty(\mathrm{SU}(2))$, i.e., this restriction on $\beta$ is irrelevant for the application we are considering.

Let us consider $\phi\in L_{2,0(\beta,\delta)}(\mathrm{SU}(2))$ such that $\Delta\phi,\Delta_P\phi\in L_{2,0(\beta,\delta)}(\mathrm{SU}(2))$ and $\norm\phi_{W^2_2(\mathrm{SU}(2))}=1$. Then, we observe
\begin{align}
  \begin{aligned}
    \norm{\Delta\phi-\Delta_P\phi}_{L_2}\le&\norm{(\Delta\phi^\downarrow)^\downarrow-(\Delta_P\phi^\downarrow)^\downarrow}_{L_2}\\
    &+\norm{(\Delta\phi^\uparrow)^\downarrow-(\Delta_P\phi^\uparrow)^\downarrow}_{L_2}\\
    &+\underbrace{\norm{(\Delta\phi)^\uparrow-(\Delta_P\phi)^\uparrow}_{L_2}}_{\le\mathrm{dim}(\mathrm{SU}(2))\delta(n)}\\
    \le&\norm{\Delta|_n\phi^\downarrow-\Delta_P|_n\phi^\downarrow}_{L_2}\\
    &+\underbrace{\norm{\Delta\phi^\uparrow-\Delta_P\phi^\uparrow}_{L_2}}_{\le\mathrm{dim}(\mathrm{SU}(2))\delta(n)}+3\delta(n)\\
    \le&\norm{\Delta|_n\phi^\downarrow-\Delta_P|_n\phi^\downarrow}_{L_2}+6\delta(n).
  \end{aligned}
\end{align}
Using $\delta(n)\to0$ and the finite order gap convergence result, we conclude that for every $\eps>0$ there exists an $n\in\mathbb{N}$ and a partitioning $P_0$ such that every partitioning $P\succeq P_0$ such that
\begin{align}
  \delta(n)<\frac{\eps}{12}.
\end{align}
Subsequently, we can find a partitioning $P_0$ such that every partitioning $P\succeq P_0$ satisfies
\begin{align}
  \norm{\Delta|_n-\Delta_P|_n}_{L(W_2^2(\mathrm{SU}(2))|_n,L_2(\mathrm{SU}(2))|_n)}<\frac{\eps}{2}.
\end{align}
With these choices of $n$ and $P_0$, we obtain for $P\succeq P_0$ (note $\norm{\phi^\downarrow}_{W_2^2(\mathrm{SU}(2))}\le\norm\phi_{W^2_2(\mathrm{SU}(2))}=1$)
\begin{align}
  \begin{aligned}
    \norm{\Delta\phi-\Delta_P\phi}_{L_2}\le&\norm{\Delta|_n\phi^\downarrow-\Delta_P|_n\phi^\downarrow}_{L_2}+6\delta(n)\\
    \le&\frac{\eps}{2}\norm{\phi^\downarrow}_{W_2^2}+6\frac{\eps}{12}\\
    \le&\eps.
  \end{aligned}
\end{align}
Hence,
\begin{align}
  \norm{\Delta-\Delta_P}_{L(W_{2,0(\beta,\delta)}^2(\mathrm{SU}(2)),L_{2,0(\beta,\delta)}(\mathrm{SU}(2)))}<\eps
\end{align}
shows norm convergence and therefore gap convergence of $\Delta_P\to\Delta$ as operators from $W_{2,0(\beta,\delta)}^2(\mathrm{SU}(2))$ to $L_{2,0(\beta,\delta)}(\mathrm{SU}(2))$.

Since gap convergence is equivalent convergence in norm-resolvent sense, we obtain that for every $\rho$ in the resolvent set $\mathcal{R}(\Delta)$ of $\Delta$, there exists $P_0$ such that for every $P\succeq P_0$ we also have $\rho\in\mathcal{R}(\Delta_P)$ and
\begin{align}
  \norm{(\rho-\Delta_P)^{-1}-(\rho-\Delta)^{-1}}_{L(L_{2,0(\beta,\delta)},W_{2,0(\beta,\delta)}^2)}\to0.
\end{align}
Finally, since the embedding
\begin{align}
  \iota:\ W_2^2(\mathrm{SU}(2))\hookrightarrow L_2(\mathrm{SU}(2))
\end{align}
is bounded with $\norm\iota_{L(W_2^2(\mathrm{SU}(2)),L_2(\mathrm{SU}(2)))}\le1$, we know that the resolvents $(\rho-\Delta_P)^{-1}$ and $(\rho-\Delta)^{-1}$ map into $L_{2,0(\beta,\delta)}$ as well and, as maps from $L_{2,0(\beta,\delta)}$ to $L_{2,0(\beta,\delta)}$, can they can be expressed as 
\begin{align}
  \iota\circ(\rho-\Delta_P)^{-1}\text{ and }\iota\circ(\rho-\Delta)^{-1}.
\end{align}
In other words, 
\begin{align}
  \norm{(\rho-\Delta_P)^{-1}-(\rho-\Delta)^{-1}}_{L(L_{2,0(\beta,\delta)},L_{2,0(\beta,\delta)})}
\end{align}
is bounded by 
\begin{align}
  \norm{(\rho-\Delta_P)^{-1}-(\rho-\Delta)^{-1}}_{L(L_{2,0(\beta,\delta)},W_{2,0(\beta,\delta)}^2)}
\end{align}
which directly implies convergence in norm-resolvent sense for the discretised Laplace in $L_{2,0(\beta,\delta)}(\mathrm{SU}(2))$.

As a final note, while $L_2(\mathrm{SU}(2))$ can be written in terms of an inductive limit of $L_{2,0(\beta,\delta)}(\mathrm{SU}(2))$ with an increasing sequence of decay functions $\delta$, the convergence does not seem to lift to the inductive limit because $\phi_n\to\phi$ in $L_2(\mathrm{SU}(2))$ does not seem to imply that $\phi_n$ has to eventually be in the same $L_{2,0(\beta,\delta)}(\mathrm{SU}(2))$ as the limit $\phi$. In fact, it is highly improbable that the discretised Laplace can converge in all of $L_2(\mathrm{SU}(2))$ in norm resolvent sense because the finite step size in the definition of $\nabla_P$ should allow for highly oscillating functions with $\nabla_P\phi=0$ but $\norm{\nabla\phi}=1$. Thus, $\Delta_P$ cannot be norm convergent to $\Delta$ as a map from all of $W_2^2(\mathrm{SU}(2))$ to $L_2(\mathrm{SU}(2))$, hence it cannot be convergent in norm-resolvent sense, and thus the embedding argument with $\iota$ fails. 

   \section{Supplementary Results}
\label{sec:all_partitionings}

While it was previously shown that the proposed canonical momentum
operators, as well as the Laplace-Beltrami operator, converge to the
correct continuum theory, the exact convergence behaviour seems
to depend on the chosen partitioning. However, gaining a full and detailed
understanding of how this choice affects e.g.\ the convergence rates
of our observables, and which features of a given partitioning
are desirable, proved to be rather difficult and ultimately beyond
the scope of this paper.

Nevertheless we would like to share our results on the topic.
Thus in this appendix, we will present the results obtained for some
of the partitionings discarded earlier, such as the original and
optimised Fibonacci and the RSC partitionings. Additionally, we
will look at two more randomly generated partitionings.

The first additional partitioning is comprised of points which are
distributed random uniformly (RU) on $S_3$. We generate these points
by generating $N$ vectors $\vec x \in\mathbb{R}^4$. First every
element of $x$ is drawn from a standard normal
distribution. Thereafter, the vector is normalised to unit length.
While these RU points are distributed uniformly on average, they are
known to cluster locally.

The next partitioning is, therefore, generated from a given RU
partitioning by maximising the distance between all next
neighbours. We denote this partitioning as distance optimised random
uniform (DoRU) partitionings. Given a RU partitioning, we numerically
maximise the sum of the squared next-neighbour distances
\begin{equation}
  \label{eq:squared-average-distance-neighbours}
  y = \sum_{\langle i,j\rangle} \lVert \vec{x_i} - \vec{x_j} \rVert^2
\end{equation}
with $\langle i,j\rangle$ denoting next neighbour pairs.

\subsection{Results}

To evaluate the additional partitionings we repeat the same tests as done previously.
The remaining data on the volume convergence is plotted in \cref{fig:volConvergenceAPDX},
the test of the commutation relations can be found \cref{fig:commConvergenceAPDX}
and the spectrum and operator convergence of the Laplace-Beltrami operator in
\cref{fig:specConvergenceLogAPDX} and \cref{fig:gapConvergenceAPDX}, respectively.

\begin{figure}
    \centering
    \includegraphics[width=0.95\linewidth]{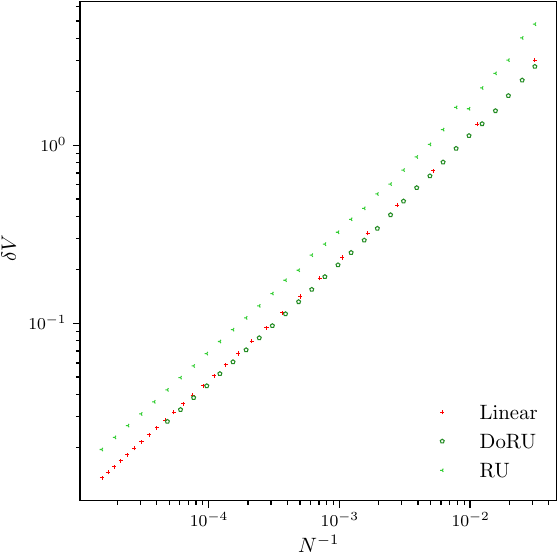}
    \caption{Like~\cref{fig:volConvergenceV1}, but comparing Linear, Distance optimised (DoRU) and Random Uniform (RU) partitionings.}
    \label{fig:volConvergenceAPDX}
\end{figure}

\begin{figure}
    \centering
    \includegraphics[width=.95\linewidth]{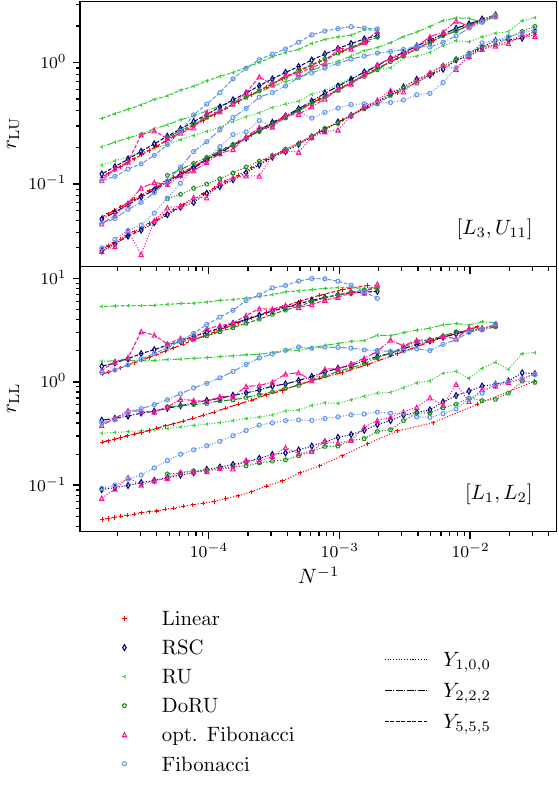}
    \vspace{-0.5cm}
    \caption{Showing the commutator convergence like in~\cref{fig:commConvergence}, comparing RSC, Distance optimised (DoRU), Random Uniform (RU) as well as optimised and unoptimised Fibonacci partitionings.}
    \label{fig:commConvergenceAPDX}
\end{figure}

\begin{figure}
  \centering
  \vspace{0.25cm}
  \includegraphics[width=\linewidth]{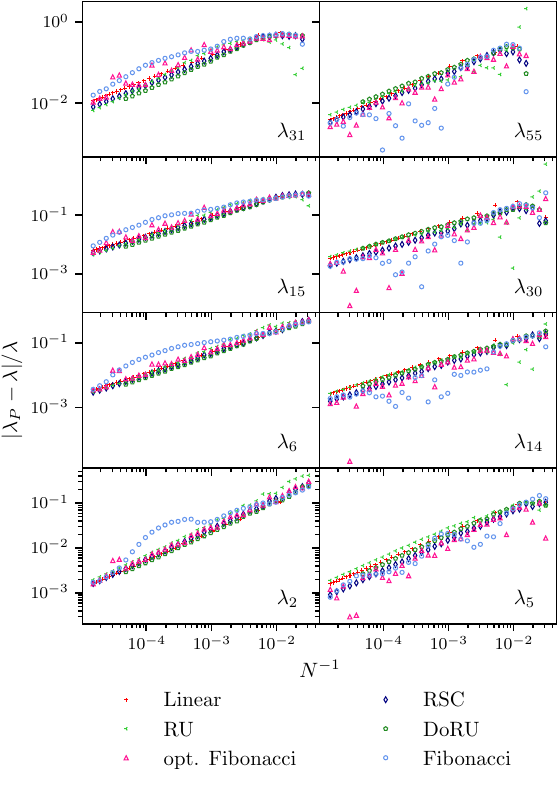}
  \vspace{-0.5cm}
  \caption{Convergence of the eigenvalues like in~\cref{fig:specConvergenceLog}, comparing RSC, Distance optimised (DoRU), Random Uniform (RU) as well as optimised and unoptimised Fibonacci partitionings.}
  \label{fig:specConvergenceLogAPDX}
\end{figure}

\begin{figure}
    \centering
    \includegraphics[width=0.95\linewidth]{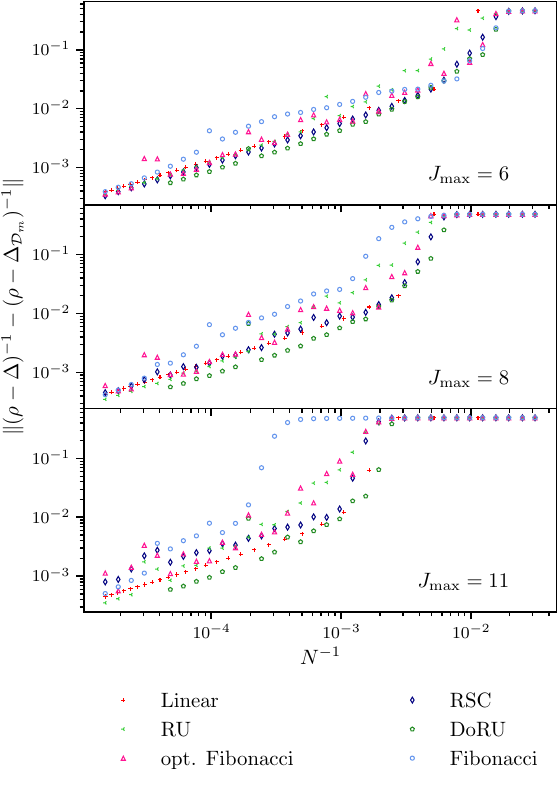}
    \vspace{-0.5cm}
    \caption{Operator convergence like in~\cref{fig:gapConvergence}, comparing RSC, Distance optimised (DoRU), Random Uniform (RU) as well as optimised and unoptimised Fibonacci partitionings.}
    \label{fig:gapConvergenceAPDX}
\end{figure}

The first thing worth noting is that the deficiencies of the unoptimised Fibonacci
lattices carry over to all the discussed observables. Bumps similar to the ones
observed in \cref{fig:volConvergenceV1} can also be found in the commutators,
spectrum and operator convergence. Optimising the coefficients $a_1$ and $a_2$
again fixes this mostly and leads to a more consistent convergence behaviour. The
more noisy convergence behaviour, however, remains, with occasional outliers visible.

As one might expect from the initial volume convergence results found in
\cref{fig:volConvergenceV2}, the RSC partitionings perform fairly similar
to the RFCC partitionings in the other observables, too.

For the RU and DoRU partitionings the situation is a bit more
intricate: in the volume, \cref{fig:volConvergenceAPDX}, RU
shows larger deviations at fixed $N$, which is a result of the local
clustering. But the convergence rate appears to be identical to the
other shown partitionings. This is also true for all the other
observables with the notable exception of $r_\mathrm{LL}$, where RU
seem to show a slower convergence, similar to the ones observed for
the Genz points. The DoRU partitioning, although initially in line
with the RSC points, also seems to tend towards this slower
convergence rate for larger $N$. At this point it is, however, unclear
whether this is inherent to the partitioning or related to the fact
that the optimisation with large $N$ becomes more and more difficult.
One more interesting result is that
the DoRU appears to have the smallest amplitude in the gap
convergence, see \cref{fig:gapConvergenceAPDX}.

 }
\end{document}